\documentclass[aps,twocolumn,nofootinbib,prd]{revtex4}
\usepackage{nicefrac}
\usepackage{latexsym}
\usepackage{amsmath, mathrsfs, bm}
\usepackage{eqnarray}
\usepackage{url}
\usepackage{cases}
\usepackage{epsfig}
\usepackage{color}
\usepackage{graphicx,subfigure}
\usepackage{grffile}
\usepackage{float}
\usepackage{natbib}
\usepackage{soul}
\usepackage{newtxtext,newtxmath}
\usepackage{multirow}

\usepackage{etoolbox}
\preto\align{\par\nobreak\small\noindent}
\expandafter\preto\csname align*\endcsname{\par\nobreak\small\noindent}
\preto\multline{\par\nobreak\small\noindent}
\expandafter\preto\csname align*\endcsname{\par\nobreak\small\noindent}
\preto\flalign{\par\nobreak\small\noindent}
\expandafter\preto\csname align*\endcsname{\par\nobreak\small\noindent}
\preto\eqnarray{\par\nobreak\small\noindent}
\expandafter\preto\csname align*\endcsname{\par\nobreak\small\noindent}

\newcommand{\citepy}[1]{\citeauthor{#1}~(\citeyear{#1})\cite{#1}}
\newcommand{\reffg}[1]{Fig.~\ref{#1}}
\newcommand{\reftb}[1]{Table~\ref{#1}}
\newcommand{\refeq}[1]{Eq.~(\ref{#1})}
\newcommand{\refsc}[1]{Sect.~(\ref{#1})}

\newcommand*{\msk}{\\[0.25cm]} 
\newcommand*{\nmsk}{\notag\msk} 

\newcolumntype{L}[1]{>{\raggedright\arraybackslash}p{#1}}
\newcolumntype{C}[1]{>{\centering\arraybackslash}p{#1}}  
\newcolumntype{R}[1]{>{\raggedleft\arraybackslash}p{#1}}

\def\planck{{\it Planck~}}

\newcommand{\minus}{\raisebox{0.75pt}{-}}

\def\rr{{\bf \hat{r}}}
\def\rv{{\bf r}}

\begin{document}

\title{Reconstructing the radial velocity profile of cosmic voids with kinematic Sunyaev-Zeldovich Effect}
\author{Yi-Chao Li$^{1,2}$}
\author{Yin-Zhe Ma$^{2,3}$}
\email{Ma@ukzn.ac.za}
\author{Seshadri Nadathur$^{4}$}
\affiliation{$^{1}$Centre for Radio Cosmology (CRC), Department of Physics and Astronomy, University of the Western Cape, Modderdam Road, Bellville, Cape Town 7530, South Africa}
\affiliation{$^{2}$School of Chemistry and Physics, University of KwaZulu-Natal, Westville Campus, Private Bag X54001, Durban 4000, South Africa}
\affiliation{$^{3}$NAOC-UKZN Computational Astrophysics Centre (NUCAC), University of KwaZulu-Natal, Durban, 4000, South Africa}
\affiliation{$^{4}$Institute of Cosmology and Gravitation, University of Portsmouth, Burnaby Road, Portsmouth, PO1 3FX, UK}


\begin{abstract}
    We develop an estimator to extract the mean radial velocity profile of cosmic 
    voids via the kinematic Sunyaev-Zeldovich effect of pairs of galaxies surrounding
    them. The estimator is tested with simulated pure kSZ map and void catalogue data 
    from the same simulation.
    The results show that the recovered signal could be attenuated by low angular
    resolution of the map or large aperture photometry filter radius size, 
    but the mean radial velocity profile can be fully recovered with our estimator. 
    By applying the estimator to the \planck {\tt 2D-ILC} CMB map, with galaxy and 
    void catalogues from BOSS DR12, 
    we find that the estimated void velocity profile is $3.31\sigma$ apart from null
    detection for for voids with continuously rising density profiles asymptoting 
    to the mean density; and $1.75\sigma$ for voids with positive density 
    contrast shell surrounded.
    By fitting the reconstructed 
    data to the theoretical profile, we find the reduced $\chi^{2}$ to be $1.19$ 
    and $0.62$ for the two types of void, respectively, indicating a good fit 
    of the model to the data.
    We then forecast the detectability of the radial velocity profile of cosmic voids 
    with future CMB surveys, including SPT-3G, AdvACT, and Simons Observatory. 
    We find that the contamination effect from CMB residuals is negligible with 
    survey area over $2,000\deg^2$, especially with aperture photometry size 
    less than $1\,{\rm arcmin}$. 
    But the effect from instrumental noise is non-negligible. For future SPT-3G, 
    AdvACT and Simons Observatory, the detection is potentially achievable 
    from $3\sigma$ to $10\sigma$ C.L., depending on specific instrumental parameters. 
    This opens a new window of probing dynamics of the cosmic structures from the 
    kinematic Sunyaev-Zeldovich effect.
\end{abstract}

\maketitle

\section{Introduction}

The kinematic Sunyaev-Zeldovich effect (hereafter, kSZ effect) is the secondary CMB anisotropy effect due to its scattering off moving electrons in the Universe~\cite{1972CoASP...4..173S,Sunyaev80}
\begin{align}
    \frac{\Delta T}{T} = -\frac{\sigma_{\rm T}}{c} 
    \int {\rm d}l\, n_{\rm e}\,(\mathbf{v}\cdot \hat{\mathbf{n}}),
\end{align}
in which, $\sigma_{\rm T}$ is the Thomson cross-section, $n_{\rm e}$ is the electron
density and $\mathbf{v}$ is the peculiar velocity of electrons relative to the CMB.
The kSZ signal was first detected by Ref.~\cite{Hand12} via
pairwise kSZ estimator proposed in Refs.~\cite{Ferreira99,Juszkiewicz00}, using the CMB data from
Atacama Cosmology Telescope (ACT~\cite{2011ApJS..194...41S}) and
the galaxy catalogue from the Sloan Digital Sky Survey III DR9~\cite{2012ApJS..203...21A}. Since then, there has been a number of works to detect the kSZ effect at various levels by cross-correlating CMB data with other large-scale structure tracers. These include the pairwise kSZ estimates with South Pole Telescope (SPT) data and Dark Energy Survey (DES) cross-correlation~\cite{Soergel16}, Atacama Cosmology Telescope (ACT) data and SDSS/DR11 cross-correlation~\cite{Bernardis17}, {\it Planck} and SDSS/DR12 cross-correlation~\cite{Sugiyama18,Li18}; kSZ cross-correlation with reconstructed peculiar velocity field~\cite{Planck-unbound,CHM15}; the squared kSZ field cross-correlation with projected density field from WISE infra-red survey~\cite{Hill16,Ferraro16}; the kSZ dispersion measurement with {\it Planck} data and X-rays selected clusters~\cite{Planck17-dispersion}. More recently, Ref.~\cite{Baxter19} detected the rotational kSZ effect with {\it Planck} map and six rotational cluster samples from SDSS/DR10.

Despite the rapid progress of kSZ detection in this field, little work has been done to investigate the behaviour of cosmic void, which is believed to dominate the cosmic volume on large scales. The cosmic voids refers to the large underdensity region of matter
distribution in the Universe. During the cosmic evolution, matters are accredited into the galaxies and clusters due to the gravitational attraction, therefore leaving more and more empty and underdense region in the Universe. The properties of the voids, such as the
abundance or the spatial distribution, are related to the initial conditions
and the universe evolution history~\cite{2004MNRAS.350..517S,2009arXiv0911.1829M}. 
In the meanwhile, the voids properties also depend on the nature of the gravity
\cite{2013MNRAS.431..749C,2015MNRAS.451.1036C}.
With $N$-body simulations, one can explore the gravitational physics of the void formation 
\cite{1993ApJ...410..458D,2001MNRAS.323....1S}, the statistical properties of the void
distribution~\cite{2009arXiv0911.1829M,2017MNRAS.467.3424T},
and the density and velocity profiles~\cite{Hamaus14,Cautun16,Demchenko16,2018arXiv181103132M}.

Recently, several cosmic void catalogues have been compiled from galaxy redshift survey data
\cite{2012MNRAS.421..926P,2012ApJ...761...44S,2014MNRAS.440.1248N,2017ApJ...835..160M}.
Studies with such cosmic void catalogue and numerical simulations suggest
that it is possible to classify voids into two different types according to their surrounding density profiles. 
The ``S-type'' voids are surrounded by a shell of positive density contrast $\delta$
which compensates the mass deficit within the void. The ``R-type'' voids instead have
continuously rising density profiles asymptoting to the mean density 
\cite{2013MNRAS.434.1435C}. The different gravitational lensing effects on the
CMB of these types of voids were recently detected \cite{Raghunathan:2020}.
The different density environment
leads to different velocity fields around cosmic voids.
Using mock galaxy and voids catalogue, \citet{2013MNRAS.434.1435C} showed that S-type 
voids have infall velocities outside of the overdense shell whereas inside 
the overdense shell the voids are expanding. In contrast, R-type voids only
expand and experience no contraction around void radius. 
Void velocity and density profiles can be measured observationally via the RSD effect 
in the void-galaxy cross-correlation \cite{2013MNRAS.436.3480P,Cai16,Pisani19,2019PhRvD.100b3504N}
or through gravitational lensing by voids \cite{Cai17,Raghunathan:2020}.

In this paper, we investigate the radial velocity profile from the measurement of 
kSZ effect, and present the current measurement from {\it Planck} data and forecasts
for future experiments. Our approach aims to provide a complementary method to 
understand the dynamical behaviour of cosmic voids than previous studies from $N$-body 
simulations, weak lensing and RSD effect. In Sec.~\ref{sec:voids}, we present the 
different types of the voids in the large-scale structure, and the estimator to 
extract the radial velocity profile of voids from kSZ effect. 
In Sec.~\ref{sec:simulation}, we calculate the expected radial velocity profile
from numerical simulations. In Sec.~\ref{sec:planck}, we present the current 
measurement of the radial velocity profile from {\it Planck} maps and BOSS DR12 
void catalogues. In Sec.~\ref{sec:forecasts}, the forecasts for future CMB 
experiments are presented and discussed, including ACT, SPT and Simons Observatory. 
The conclusion is presented in the last section. 

Throughout this paper, we adopt a spatially flat $\Lambda$-Cold-Dark-Matter
($\Lambda$CDM) cosmology model with cosmological parameters fixed at {\it Planck}
2015 best-fitting values, i.e. 
$\Omega_{\rm b} h^2 = 0.02230$, $\Omega_{\rm c} h^2 = 0.1188$, 
$\Omega_{\rm k} h^2 = 0.00037$ and 
$H_0=67.74\,{\rm km}\,{\rm s}^{-1}\,{\rm Mpc}^{-1}$ \cite{2016A&A...594A..13P}.

\section{The cosmic voids}
\label{sec:voids}
\subsection{Two types of cosmic voids}
It is usefull to characterise the population of voids into R-type and S-type
\cite{2004MNRAS.350..517S,Cai14,Demchenko16,Cautun16,Pisani19}. 
The S-type voids usually lie within some overdense region of large-scale structure, 
therefore it is usually called ``Void-in-cloud'' scenario. 
In comparison, R-type of voids is usually embedded in the low-density region of 
large-scale structure, and therefore named as "Void-in-void".

Ref.~\cite{2017MNRAS.467.4067N} investigated the gravitation potential environments
of the voids and suggested a empirical linear relation between the 
averaged value of potential and the void properties,
\begin{align}
    \bar{\Phi}_0 = -a \lambda_v + c,
\end{align}
in which, $a$ and $c$ are positive constants determined from the simulation
and the parameter $\lambda_v$ is defined as
\begin{align}
    \lambda_v = \bar{\delta}_{\rm g} \left(\frac{R_{\rm eff}}{h^{-1}{\rm Mpc}}\right)^{1.2}
\end{align}
where $R_{\rm eff}$ is the effective void size and $\bar{\delta}_{\rm g}$ is the 
average galaxy density contrast over the void. The linear scaling of 
$\bar{\Phi}_0$ and $\lambda_v$ is universal and relatively independent of
the galaxy tracer properties. According to the analysis in 
\cite{2017MNRAS.467.4067N},  
voids with larger values of $\lambda_v$, particularly $\lambda_v>10$ for 
LOWZ/CMASS catalogue, are more strongly correlated
with the regions of negative gravitational potential, $\Phi<0$,
and more likely to be the S-type voids lying in the overdensity regions. In contrast, 
the smaller values of $\lambda_v$ indicate the R-type voids with underdensity 
environments. In our analysis, we split the void catalogue in to two sub-samples 
with $\lambda_v$ greater and less than $10$. The R-type of voids is likely to have a positive velocity profile pointing outward from the center of the void; while the S-type of voids is likely to have a slightly negative velocity profile at the boundary region of the voids due to the over dense ``wall'' surrounding them.
Recently, \citet{Raghunathan:2020} observationally confirmed the difference 
in void profiles with $\lambda_v$ based on their CMB lensing signal.

\subsection{The estimator of void radial velocity profile}\label{sec:estimator}

\subsubsection{The pairwise kSZ estimator}
We want to use the relative velocity between the galaxies around a cosmic void to extract the radial velocity profile of the void. For this reason, we need to calculate the normal pairwise momentum estimator between two galaxies due to the normal gravitational attraction, then subtract this term from the true pairwise kSZ effect.  The pairwise momentum estimator was initially proposed by Refs.~\cite{Ferreira99,Juszkiewicz00} and used in Hand et al. (2012)~\cite{Hand12} to make the first detection of the kSZ effect. The estimator has been widely used in previous kSZ studies~\cite{Planck-unbound,Soergel16,Bernardis17,Sugiyama18,Li18}. By selecting a pair of galaxies, $i$ and $j$,
we can measure the difference between CMB temperature fluctuations $\Delta_{ij} = T_i - T_j$.
$T_i$ and $T_j$ are the CMB temperatures where the two galaxies
locate, filtered with an aperture photometry (AP) filter or matched filter
to removed the large-scale primary CMB contribution. 
The $\chi^2$ function can be constructed as
\begin{align}
    \chi^2 = \sum_{i<j} \left[ \Delta_{ij} - \Delta^{\rm p}(r_{ij}) P_{ij} \right]^2,
\end{align}
where indices $i$ and $j$ sum over all pairs without repetition. $P_{ij}$ is the geometric operator that projects the pairwise velocity to the 
line-of-sight(LoS) direction. It can be expressed as~\cite{Ferreira99,Juszkiewicz00}
\begin{align}
    P_{ij} & = \left(\rr_{ij}\right)\cdot\frac{\left(\rr_i+\rr_j\right)}{2} \nmsk
    & = \frac{\left(r_i - r_j\right)\left(1 + \cos\theta\right)}
    {2\sqrt{r_i^2 + r_j^2 - 2r_ir_j\cos\theta}}
\end{align}
where $\rr_{i,j} = \rv_{i,j} / r_{i,j} $, $\rr_{ij} = (\rv_i - \rv_j)/r_{ij}$.
$r_{i}$ and $r_{j}$ are the comoving distance of the galaxies $i$ and $j$, respectively;
and $r_{ij}=|\mathbf{r}_{i}-\mathbf{r}_{j}|$.

$\Delta^{\rm p}(r)$ is the pairwise kSZ effect due to the mean pairwise 
velocities of the galaxies and it can be expressed with the mean CMB temperature
$T_{\rm CMB}$, the mean optical depth of the galaxies $\bar{\tau}$, speed of light $c$
and the mean pairwise velocity $v^{\rm p}(r)$ as,
\begin{align}\label{eq:T2v}
    \Delta^{\rm p}(r) = - \frac{T_{\rm CMB} \bar{\tau}}{c}v^{\rm p}(r)
\end{align}
where $r$ is the comoving separation distances between the pair of samples.

By minimizing the $\chi^2$ function, 
$\partial \chi^2 / \partial \hat{\Delta}^{\rm p} = 0$ leads to the expression of estimator~\cite{Ferreira99,Juszkiewicz00,Hand12}
\begin{align}
    \hat{\Delta}^{\rm p}(r) = \frac{\sum_{i<j} \Delta_{ij} P_{ij}}{\sum_{i<j}P_{ij}^2}.
\end{align}
We will apply the above estimator to {\it Planck} maps and simulated kSZ maps of, SPT-3G, ACT and Simons Observatory.

\subsubsection{kSZ effect due to the void expansion}

If the galaxies are near the void, they are pushed by the void in the direction away 
from its center. 
Therefore there is an outflow component to the velocity, $v^{\rm e}(R)$, 
caused by the expansion of the void, where $R$ is the radial distance relative 
to the void centre. 
We assume that each void share the same radial velocity profile normalised by the 
effective radius of the void, i.e. $v^{\rm e}(R/R_{\rm eff})$. We want to reconstruct 
this radial velocity profile by building an estimator $\hat{v}^{\rm e}(R)$ for the 
observed kSZ map.

In the comoving frame, we define the vector from observer pointing to the center of the 
void as $\mathbf{r}_{o}$, and to the selected galaxy as $\mathbf{r}_{i}$. Then the 
vector from the void center to the galaxy sample is $\mathbf{r}_{io}$. Then 
projection of $\mathbf{r}_{io}$ onto the LoS direction is
\begin{align}
    q_i = \rr_{io} \cdot \rr_i
    = \frac{r_i - r_o \cos\theta_{io} }{\sqrt{r_i^2 + r_o^2 - 2 r_i r_o \cos\theta_{io}}},
\end{align}
in which $\theta_{io}$ is the separation angle between the galaxy and the void centre.
Then the velocity difference projected onto the LoS direction becomes $v^{\rm e}(R) Q_{ij}$,
where $Q_{ij} = q_i - q_j$. Then temperature difference due to the kSZ effect
can be expressed as
\begin{align}
    \Delta^{\rm e}_R Q_{ij} = - \frac{T_{\rm CMB} \bar{\tau}}{c} v^{\rm e}(R) Q_{ij}.
\end{align}
The $\chi^2$ function can be expressed as
\begin{align}
    \chi^2 = \sum_{i<j} \left[ \Delta_{ij} - \left(
    \Delta^{\rm p}(r_{ij})P_{ij} + \Delta^{\rm e}_R Q_{ij}\right) \right]^2.
\end{align}
Taking the minimal condition as $\partial \chi^2 / \partial \Delta^{\rm e}_R = 0$, 
we obtain the estimator
\begin{align}
    \hat{\Delta}^{\rm e}_R = \frac{\sum_{i<j} \Delta_{ij}Q_{ij}}{\sum_{i<j}Q_{ij}^2}
    - b_R, \label{eq:deltaR}
\end{align}
where
\begin{align}
b_R = \frac{\sum_{i<j}\Delta^{\rm p}(r_{ij})P_{ij}Q_{ij}}{\sum_{i<j}Q_{ij}^2} \label{eq:bR}
\end{align}
is a bias term due to the normal gravitational attraction between galaxy pairs. Finally, 
the radial velocity profile of the void can be estimated as
\begin{align}
    \hat{v}^{\rm e}(R) = - \frac{c}{T_{\rm CMB} \bar{\tau}} \hat{\Delta}^{\rm e}_R. \label{eq:vR}
\end{align}

\section{Expected signal}\label{sec:simulation}

\subsection{Simulation data}

We use one redshift snapshot at $z=0.52$ from the Big MultiDark (BigMD)
$N$-body simulation \cite{2016MNRAS.457.4340K,2012MNRAS.423.3018P}, 
which follows the evolution of $3840^3$ particles in a box of side 
$L=2500\,h^{-1}{\rm Mpc}$ using \textsc{GADGET}-2\cite{2005MNRAS.364.1105S}
and adaptive refinement tree~\cite{1997ApJS..111...73K,2008arXiv0803.4343G} codes. The halo catalogues are formed by using the Bound Density Maximum algorithm 
\cite{1997astro.ph.12217K,2013AN....334..691R} and the underlying 
dark matter density field is determined from the full resolution simulation 
output on a $2350^3$ grid using the cloud-in-cell interpolation.

\begin{itemize}

    \item The {\it mock galaxy catalogues} are created by assigning galaxies to DM halo
        following a distribution based on the halo mass. The halos are populated 
        according to the halo occupation distribution (HOD) model
        \cite{2007ApJ...667..760Z}. 
        The catalogue contains central and satellite galaxies.
        In our analysis, we only use the central 
        galaxies, which are good tracers of the clusters velocities.
        The central galaxies were placed at the center
        of their respective halos and the numbers of central galaxies in each 
        mass bin follow a nearest integer distribution with the mean occupation function,
        \begin{align}
            \langle N_{\rm cen} (M) \rangle = \frac{1}{2}\left[1 + 
            {\rm erf} \left( \frac{\log M - \log M_{\rm min}}{\sigma_{\rm log M}}
            \right) \right],
        \end{align}
        in which, the parameters $M_{\rm min}$, $\sigma_{\rm log M}$ were chosen in order
        to match those the properties of the SDSS CMASS catalogue.
        For the details of mock galaxy catalogue generation, please refer to
        \citepy{2017MNRAS.467.4067N}.

    \item The {\it voids catalogues} are identified using the \texttt{REVOLVER} 
        void-finding algorithm \cite{2019PhRvD.100b3504N}, which is based on the 
        earlier \texttt{ZOBOV} 
        (ZOnes Bordering On Voidness) watershed void-finding algorithm
        \cite{2008MNRAS.386.2101N}.
        \texttt{REVOLVER} estimates the local galaxy density field from the discrete galaxy 
        distribution using a Voronoi tessellation method, which includes additional
        corrections 
        for the survey selection function and angular completeness, described in detail in
        \cite{2014MNRAS.440.1248N,2016MNRAS.461..358N}.
        The local minima in this field
        are identified as the center of the voids and the watershed basins around them
        are the void edges. Following the procedure of Ref.~\cite{2016MNRAS.461..358N},
        the voids are identified with each individual density basin without any
        additional merging.
        The size of the void is characterized by the effective void radius 
        $R_{\rm eff} = (3 V/ 4\pi)^{1/3}$, where $V$ is the total void volume 
        determined from the sum of its Voronoi cells volume.

\end{itemize}

\subsection{Simulated kSZ signal}\label{sec:simksz}

\begin{figure*}
    \small
    \centering
    \includegraphics[width=0.9\textwidth]{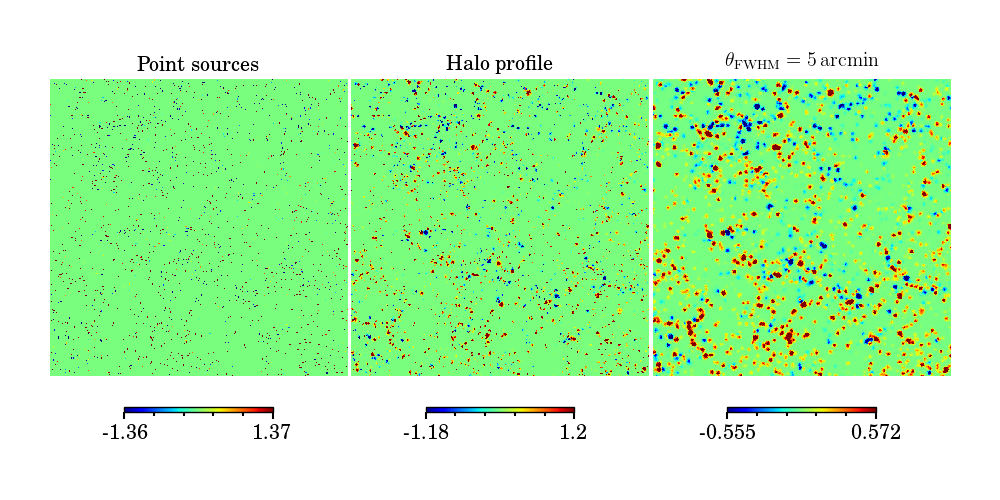}
    \caption{
        Simulated kSZ maps. 
        {\it Left}-- assuming point sources;
        {\it Middle}--assuming Gaussian halo profile for the clusters;
        {\it Right}: Gaussian halo profile map smoothed with Gaussian beam function with
        $\theta_{\rm FWHM} = 5\,$arcmin.
        The color ranges for all the plots are in unit of $\mu{\rm K}$ and
        truncated at $\pm 3\sigma$.
    }\label{fig:sim}
\end{figure*}

We use the position and velocity information of the central galaxies
to simulate the kSZ signal.
The mock galaxy catalogue is firstly split into $5$ sub-catalogues 
along the $\hat{\mathbf{z}}$-axis, which is chosen as the Line-of-Sight (LoS) direction.
Each of the sub-catalogues is assigned new radial comoving distance 
with median values equal to the comoving distance at redshift $z=0.52$.

The kSZ temperature anisotropies is given by
\begin{align}\label{eq:ksz}
    \Delta T_{\rm kSZ} = -\frac{T_{\rm CMB}\bar{\tau}}{c}v_{\rm LoS}
\end{align}
in which, $\bar{\tau}$ is a free rescaling parameter as an 
average optical depth of the central galaxy samples. We assume $\bar{\tau}=10^{-3}$
for the simulation. We will show later, that this parameter is sensitive to the choice of samples,  angular resolution of the CMB map and the filter size of aperture photometry (AP).
$v_{\rm LoS}$ denotes the LoS component of
peculiar velocity of free electrons.

At the stage-1 of simulation, we assume the clusters to be 
point sources and traced by the central galaxies. 
We project the central galaxy catalogues to the \texttt{HEALPix} maps with 
$N_{\rm side} = 2048$. The pixels with galaxies are assigned
the kSZ temperature according to \refeq{eq:ksz}. A $10^\circ\times10^\circ$
patch of the simulated map is shown in the left panel of \reffg{fig:sim}.

At the stage-2 of the simulation, we include the halo profile for the clusters.
The velocity within the clusters is assumed to be constant and the
$\bar{\tau}$ follows halo profile. To simplify the analysis, we use
Gaussian halo profile. 
\begin{align}
    \tau(r) = \bar{\tau} \exp\left[-\frac{r^2}{\sigma_{\rm vir}^2}\right],
\end{align}
in which, $r$ is the angular separation of the LoS to the halo center; 
and $\sigma_{\rm vir}$ is the angular size of the halo virial radius.
The halo virial radius is estimated according to the halo virial mass
individually~\cite{2011ApJS..192...18K},
\begin{align}
    M_{\rm vir} &= \frac{4\pi}{3}\left[\Delta_{\rm c}(z)\rho_{\rm c}(z)\right]
    r_{\rm vir} ^ 3  \nmsk
    \Delta_{\rm c}(z) &= 18\pi^2 + 82\left[ \Omega(z) - 1 \right]
    - 39\left[ \Omega(z) - 1 \right] ^ 2,
\end{align}
where $\rho_{\rm c}(z)=\rho_{\rm c0}E^{2}(z)$ is the critical density at redshift $z$. $\rho_{\rm c0}=1.879\,h^{2} \times 10^{-29}\,{\rm g}\,{\rm cm}^{-3}$, $E(z)=\sqrt{\Omega_{\rm m}(1+z)^{3}+\Omega_{\Lambda}}$, and $\Omega(z)=\Omega_{\rm m}(1+z)^{3}/E^{2}(z)$. The middle panel of \reffg{fig:sim} shows the
simulated map with halo profile applied.

At the stage-3 of the simulation, we convolve the simulated maps with Gaussian beam of
$\theta_{\rm FWHM} = 5\,{\rm arcmin}$. The smoothed patch is 
shown in the right panel of \reffg{fig:sim}.

\subsection{Radial velocity profile}
\begin{figure}
    \small
    \centering
    \includegraphics[width=0.49\textwidth]{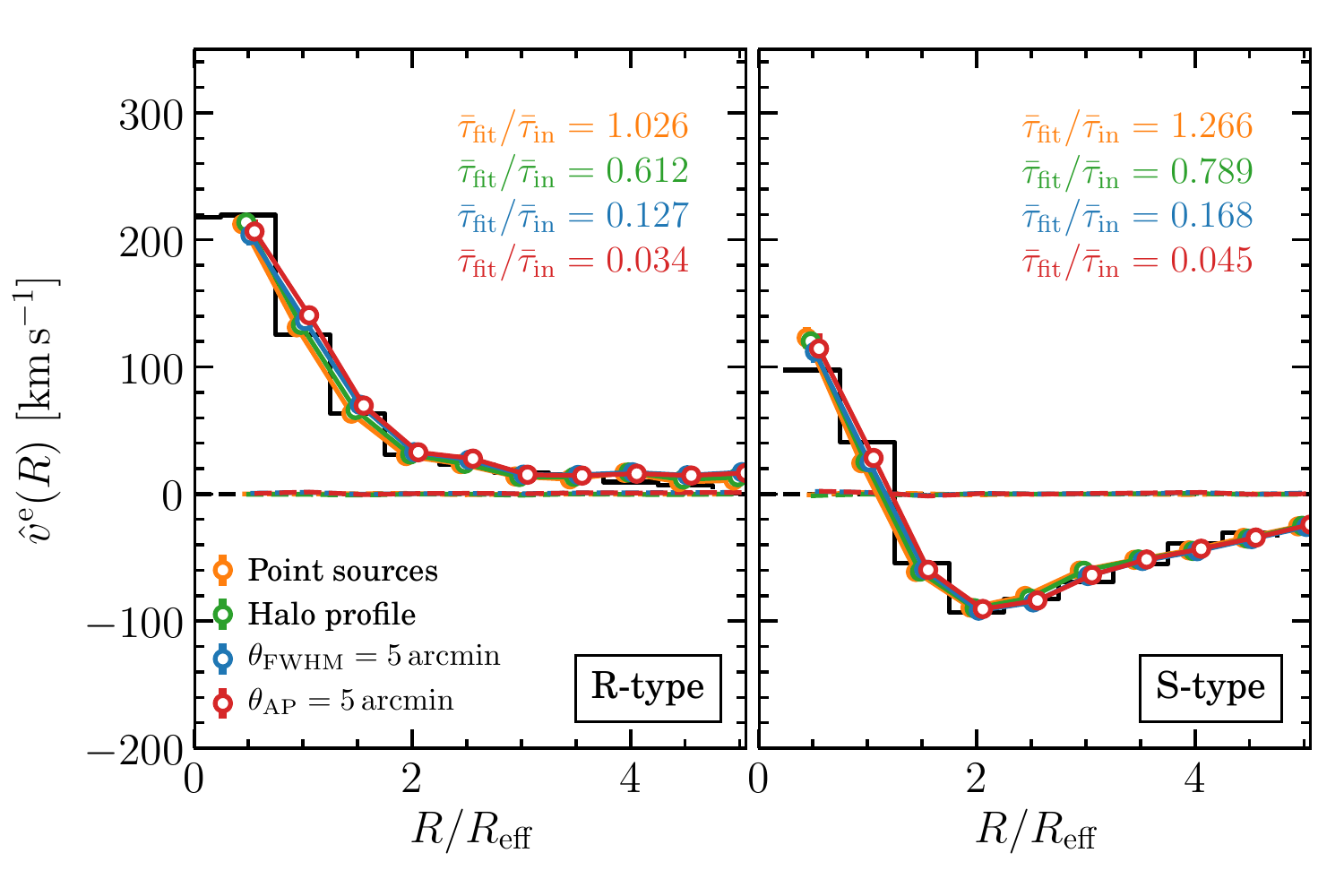}
    \caption{
        {\it Left}--the results for voids with $\lambda_v < 10$ (R-type);
        {\it Right}-- the results for voids with $\lambda_v > 10$ (S-type).
        The black step line shows the theoretical radial
        velocity profile which is directly read from the
        simulated catalogue.
        The estimate radial velocity profiles with different stages 
        of simulation maps are fit to the theoretical curve with $\bar{\tau}$
        as the free fitting parameter. The ratio of the fitted $\bar{\tau}_{\rm fit}$
        to input $\bar{\tau}_{\rm in}$ are also list, where
        $\bar{\tau}_{\rm in} \equiv 10^{-3}$. 
        The dashed lines are the mean $\hat{v}^{\rm e}$ estimated with 
        the pixels randomly-shuffled CMB temperature values
        (see \refsc{sec:results} for details).
        }\label{fig:simlv}
\end{figure}

We first calculate the radial velocity profile of cosmic voids as black step lines 
in Fig.~\ref{fig:simlv}. Since each void has different radius, we plot the radial 
velocity profile as a function of normalized radius $R/R_{\rm eff}$, where $R_{\rm eff}$
is the effective void radius. The void catalogue is split into R-type and S-type 
subsamples according the void parameter $\lambda_v$.
The left panel of \reffg{fig:simlv} shows the case with R-type voids ($\lambda_v<10$) 
and right panel shows the S-type voids ($\lambda_v>10$). 
One can see that, a significantly different radial velocity profiles between
these two types are shown. The R-type voids have a positive velocity profile
up to the radial distance of 5 times of void effective radius, indicating an overall expansion of the voids. The velocity profile
of the S-type voids is positive within the void effective radius and
negative outside, which indicates that at the boundary region the structures are 
actually collapsing. This velocity structure is consistent with the previous 
investigation of the radial density profile through numerical simulations~
\cite{2004MNRAS.350..517S,2013MNRAS.436.3480P,Demchenko16}.

We then test the behaviour of our estimator (Eqs.~(\ref{eq:deltaR})-(\ref{eq:vR})) 
with simulated kSZ maps at the three different stages as mentioned in 
Sect.~\ref{sec:simksz} and Fig.~\ref{fig:sim}. We first estimate the the voids 
expansion rate by using our estimator (Eq.~(\ref{eq:vR})) and then fit the 
estimated $\tilde{v}^{\rm e}$ 
to the theoretical prediction with $\bar{\tau}$ as the free scaling parameter.
The values of $\bar{\tau}_{\rm fit}/\bar{\tau}_{\rm in}$, which are
the ratio of best-fitting to the initial input value of
$\bar{\tau}_{\rm in} \equiv 10^{-3}$,
are list in the legend with corresponding colors
for each individual simulation stage.

The radial velocity profile estimated with the first stage simulation map,
which assumes the cluster to be point sources, is shown with orange color. 
Both the results for R-type and S-type voids are fitted to the theoretical 
predictions very well. The values of
$\bar{\tau}_{\rm fit}/\bar{\tau}_{\rm in}=1.026\pm0.04$ for 
R-type of voids, which is consistent with unity.
For S-type voids, 
$\bar{\tau}_{\rm fit}/\bar{\tau}_{\rm in}=1.266\pm0.06$, 
which indicates a larger $\bar{\tau}$ comparing to the mean values.
A possible reason is that, the S-type voids are mostly located in
overdense region, where the $\bar{\tau}$ is over the global mean values.

The results with the simulation maps considering the halo profiles are shown
as green lines; and the results with further including the beam
smoothing effect are shown as blue lines. Both of these two sets of results
are fitted to the theoretical predictions very well, except for the decreasing values of the
$\bar{\tau}_{\rm fit}/\bar{\tau}_{\rm in}$. Therefore, both Gaussian profile 
and the beam effect can smooth the signal of the void expansion effect, 
producing a lower estimated value for the mean optical depth of clusters.

In the kSZ studies with real observational data, the primary fluctuation of the 
CMB is usually the contaminant factor at the background level. 
Therefore, the aperture photometry method (AP) is usually used as a filter to 
remove the long-wavelength mode fluctuations of the CMB~\cite{CHM15,Bernardis17,Li18}. 
In the AP method, we compute the average temperature within a given angular radius 
$\theta$ and subtract from it the average temperature in a surrounding ring with 
inner and outer radii $[\theta,\sqrt{2}\theta]$. This method is model-independent 
and blinds to any specific model of spatial distribution of gas. 
As suggested by Ref.~\cite{Li18}, we use the aperture size $\theta_{\rm AP}=5\,$arcmin 
which can optimize the detection.

We apply the AP filter to our Stage-3 simulation and show the results in red lines
in Fig.~\ref{fig:simlv}. One can see that the profile can be well fitted by the
theoretical velocity profile, but the mean optical depth is reduced by almost $2$ orders of magnitude.
The signal reducing due to the smoothing effect makes the detection harder 
if systematic noise, such as the equipment noise, CMB and thermal SZ residuals are 
presented.

\section{Measurements with {\it Planck} maps}\label{sec:planck}

\begin{figure}
    \small
    \centering
    \includegraphics[width=0.45\textwidth]{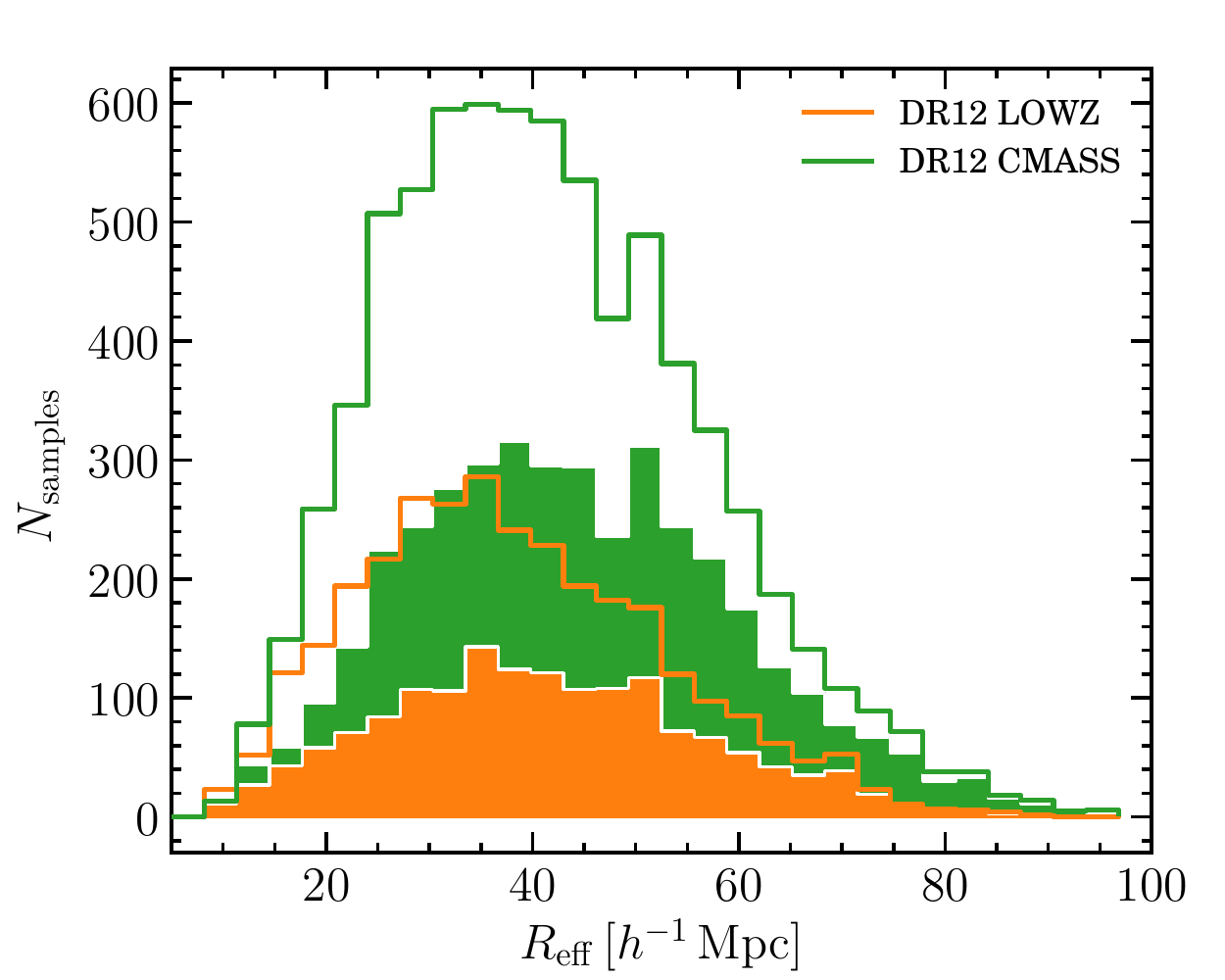}
    \caption{
        The empty histograms show the distribution of the total 
        LOWZ/CMASS (orange/green) void catalogue; The filled 
        histograms show the distribution of the void with $\lambda_v<10$
        (R-type voids). The difference between empty and filled 
        histograms is therefore for S-type voids, which have
        $\lambda_v>10$. 
    }\label{fig:vreff}
\end{figure}

\subsection{Data}
Here we present the real data in our estimation of radial velocity profile of cosmic voids.

\paragraph{\planck CMB maps}
We use two different thermal SZ-free CMB maps in our study to cross-check the consistency.
One is the {\tt 2D-ILC} CMB map,
which is obtained by applying the `Constrained ILC' component separation
method \cite{2011MNRAS.410.2481R} to the public 
\planck 2015 data\footnote{\url{http://pla.esac.esa.int/pla}}. 
This algorithm is similar to the {\it Planck} NILC method~\cite{Delabrouille09,Basak12}
which performs a minimum-variance weighted linear combination of the $9$ {\it Planck} 
frequency maps, with weights calculated to give response to CMB spectral distortion. 
In addition to this, the algorithm is designed to nullify the tSZ signal so there is 
no tSZ bias and variance in this map. The Full-Width-Half-Maximum (FWHM) of the map 
is $\theta_{\rm FWHM}=5\,$arcmin. For more details of the {\tt 2D-ILC} map, 
we refer to Refs.~\cite{2011MNRAS.410.2481R} and~\cite{Planck17-dispersion}.

As a comparison, we also use the {\tt SMICA-noSZ} map from {\it Planck} 
survey~\cite{Planck-diffuse18} in this analysis. 
The Spectral Matching Independent Component Analysis (SMICA) method produces a 
foreground-cleaned CMB map from a linear combination of multi-frequency sky maps in 
harmonic space~\cite{Cardoso08}. Similar to {\tt 2D-ILC} map, this method also imposes 
a linear constraint to nullify the frequency dependence of the tSZ. 
The angular resolution of the {\tt SMICA-noSZ} map is the same as {\tt 2D-ILC} map.

The major difference between the two methods is that the {\tt 2D-ILC} map uses 
wavelet decomposition to clean foregrounds locally both in pixel space and harmonic space,
whereas the {\it Planck} {\tt SMICA-noSZ} map is performed only in harmonic space. 
In addition, the {\tt 2D-ILC} map was produced based on {\it Planck} 2015 data release, 
while the {\tt SMICA-noSZ} map is based on {\it Planck} 2018 data release. 
Therefore, {\tt SMICA-noSZ} map is less noisy and also has some improvements 
on calibrations of {\it Planck} instrumental systematics.

\paragraph{Central Galaxy Catalogue}
We adopt the galaxy catalogue
of the twelfth data release of the Baryon Oscillation Spectroscopic Survey
(BOSS DR12) and further select the ``Central Galaxies'' from the catalogue. 
The BOSS sample is designed to measure the BAO signature 
in two-point correlation function of the galaxy clustering. The BOSS/DR12 
catalogue is separated into ``LOWZ'' and ``CMASS'' catalogues. 
Both catalogues include two separated survey areas, located in Northern and 
Southern Galactic Caps (NGC and SGC). In our analysis, we combine NGC and SGC
samples.

The ``LOWZ'' samples span the redshift below $0.45$; and
the ``CMASS'' samples span the redshift between $0.45$ and $0.8$ 
~\cite{2016MNRAS.455.1553R}. We cut off samples with $z<0.01$.
We further reduce the catalogue by choosing the ``Central Galaxies'' which are 
isolated and dominant galaxies with no other galaxies within 
$1.0\,{\rm Mpc}$ in the transverse direction and redshift difference 
smaller than $1,000\,{\rm km\,s}^{-1}$~\cite{2016A&A...586A.140P,Li18}. 
The selected central galaxies samples are regarded as the good indicator of the 
gravity center and therefore represent the clusters' center positions. 
We refer the reader to Ref.~\cite{Li18} for more details of the selection process.

\paragraph{Void catalogue}
The void catalogue used in our analysis are also generated with the 
BOSS DR12 LOWZ/CMASS LSS galaxy samples~\cite{2016MNRAS.461..358N} and
identified via \texttt{REVOLVER} void-finding algorithm, which is the 
same method as we used for our simulation void catalogue.
The void size distribution of the LOWZ/CMASS void catalogue are shown
in \reffg{fig:vreff}. The empty histograms in \reffg{fig:vreff}
show the void size distribution of the total void catalogue, 
with orange color for LOWZ catalogue and green color for CMASS catalogue,
respectively. The size distribution of the R-type voids, which are void
with $\lambda_v<10$ is shown with the filled histogram and the 
differences between empty and filled histogram show the distribution 
of S-type voids. One can see that for both R-type and S-type voids in both 
LOWZ and CMASS catalogues, the peak distribution of the void size is around 
$R_{\rm eff}\simeq 40\,h^{-1}{\rm Mpc}$. The total number of R-type and S-type 
voids for LOWZ are $1,462$ and $1,348$ and for CMASS are 
$3,845$ and $3,122$, shown in Table~\ref{tab:chi2-real}. 
This indicates that CMASS volume contains more voids than LOWZ volume, 
and within each survey volume there is similar number of R-type and S-type voids.
There's no significant difference in voids size distribution between 
R-type and S-type voids.

\subsection{Results of computation}\label{sec:results}

\begin{figure*}
    \small
    \centering
    \begin{minipage}[t]{0.49\textwidth}
        \includegraphics[width=\textwidth]{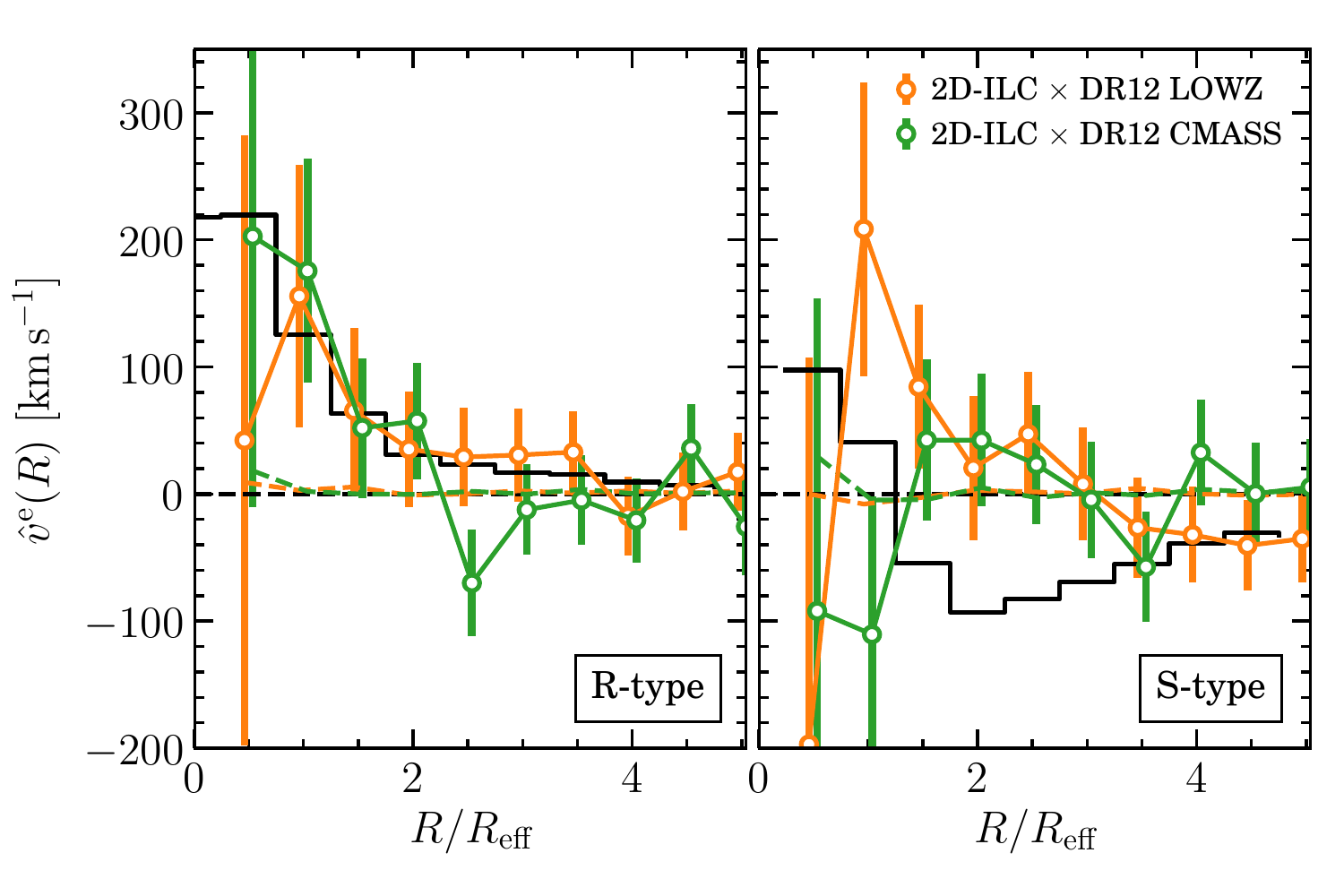}
        \caption{
            The estimated $\hat{v}^{\rm e}(R)$ with \planck {\tt 2D-ILC}
            CMB map and LOWZ/CMASS (orange/green) void
            catalogue. The black step line shows the theoretical 
            prediction which is directly read from the
            simulated catalogue.
            The {\it left} and {\it right} panels show the results for R-type and S-type
            voids, respectively. The estimated kSZ temperature difference
            are converted to velocity using \refeq{eq:T2v}. 
            We use the $\bar{\tau}$ fitted with R-type results for both
            R-type and S-type temperature-velocity conversion. The fitted
            $\bar{\tau}$ values are summarized in \reftb{tab:chi2-real}.
        }\label{fig:dr12lv}
    \end{minipage}\hfill
    \begin{minipage}[t]{0.49\textwidth}
        \includegraphics[width=\textwidth]{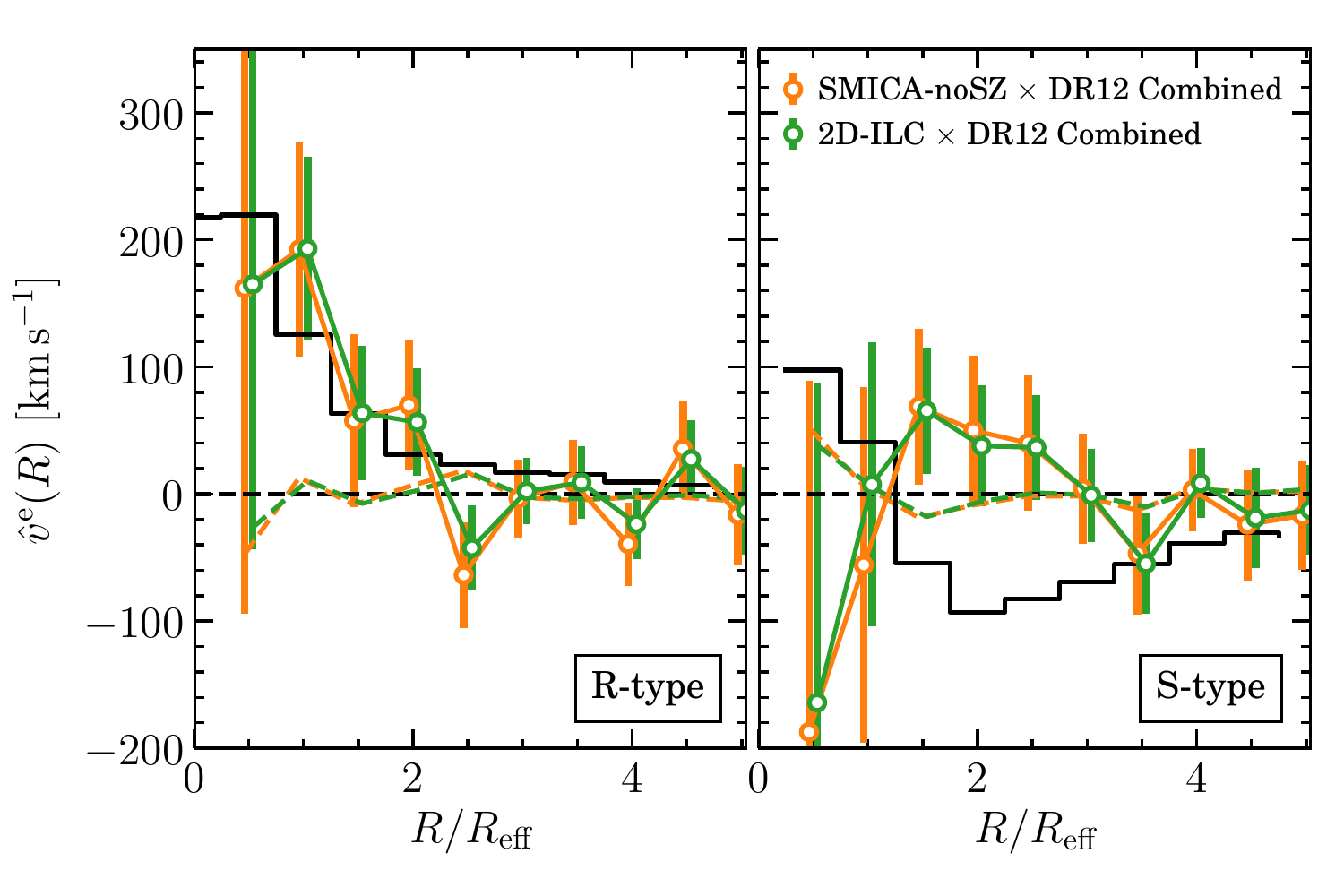}
        \caption{
            Same as \reffg{fig:dr12lv}, but using 
            \planck {\tt SMICA-noSZ}/{\tt 2D-ILC} maps (orange/green)
            and combined LOWZ and CMASS catalogue. 
            We use the $\bar{\tau}$ fitted with R-type results for both
            R-type and S-type temperature-velocity conversion. The fitted
            $\bar{\tau}$ values are summarized in \reftb{tab:chi2-real}.
        }\label{fig:SMICAlv}
    \end{minipage}
\end{figure*}

\begin{figure*}
    \small
    \centering
    \includegraphics[width=0.49\textwidth]{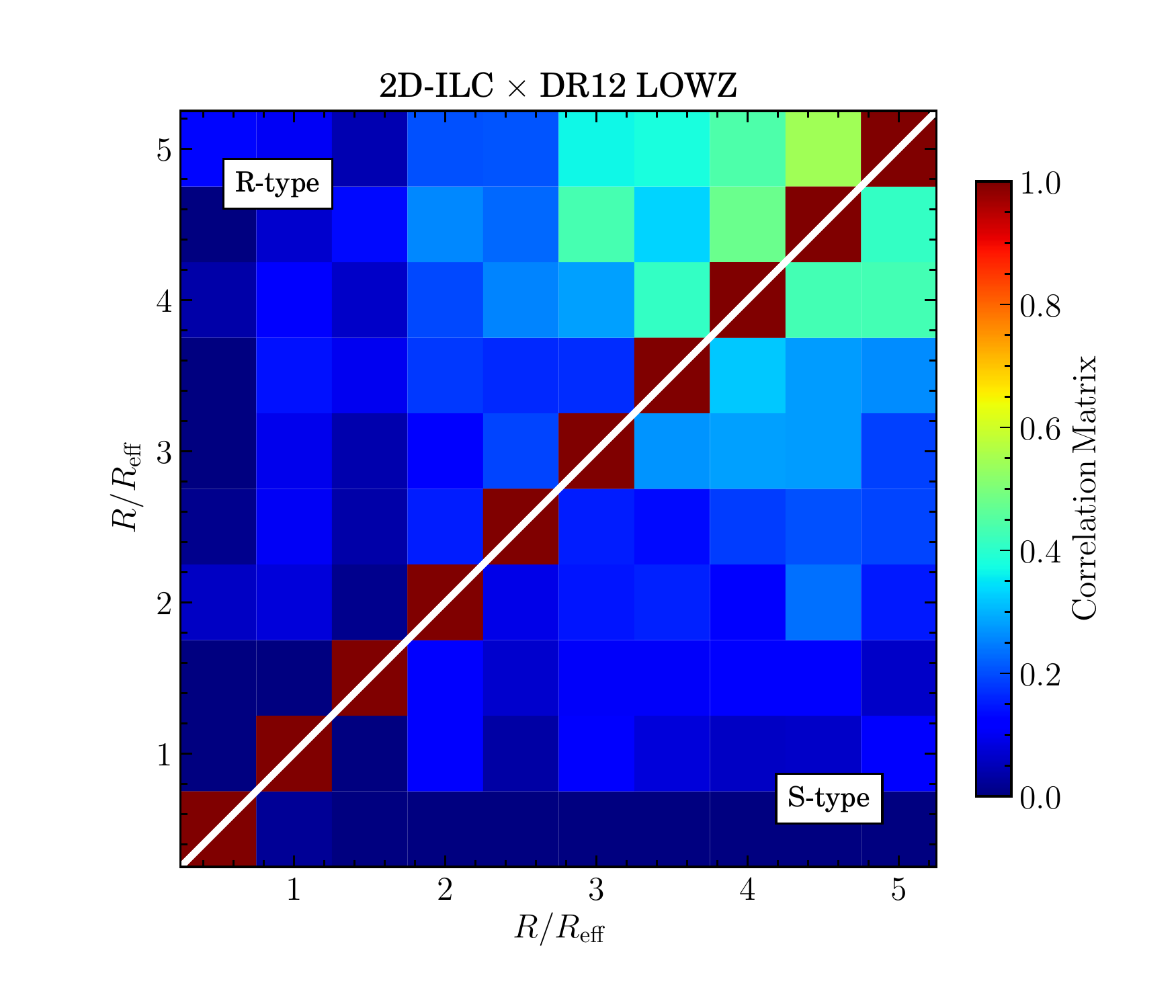}
    \includegraphics[width=0.49\textwidth]{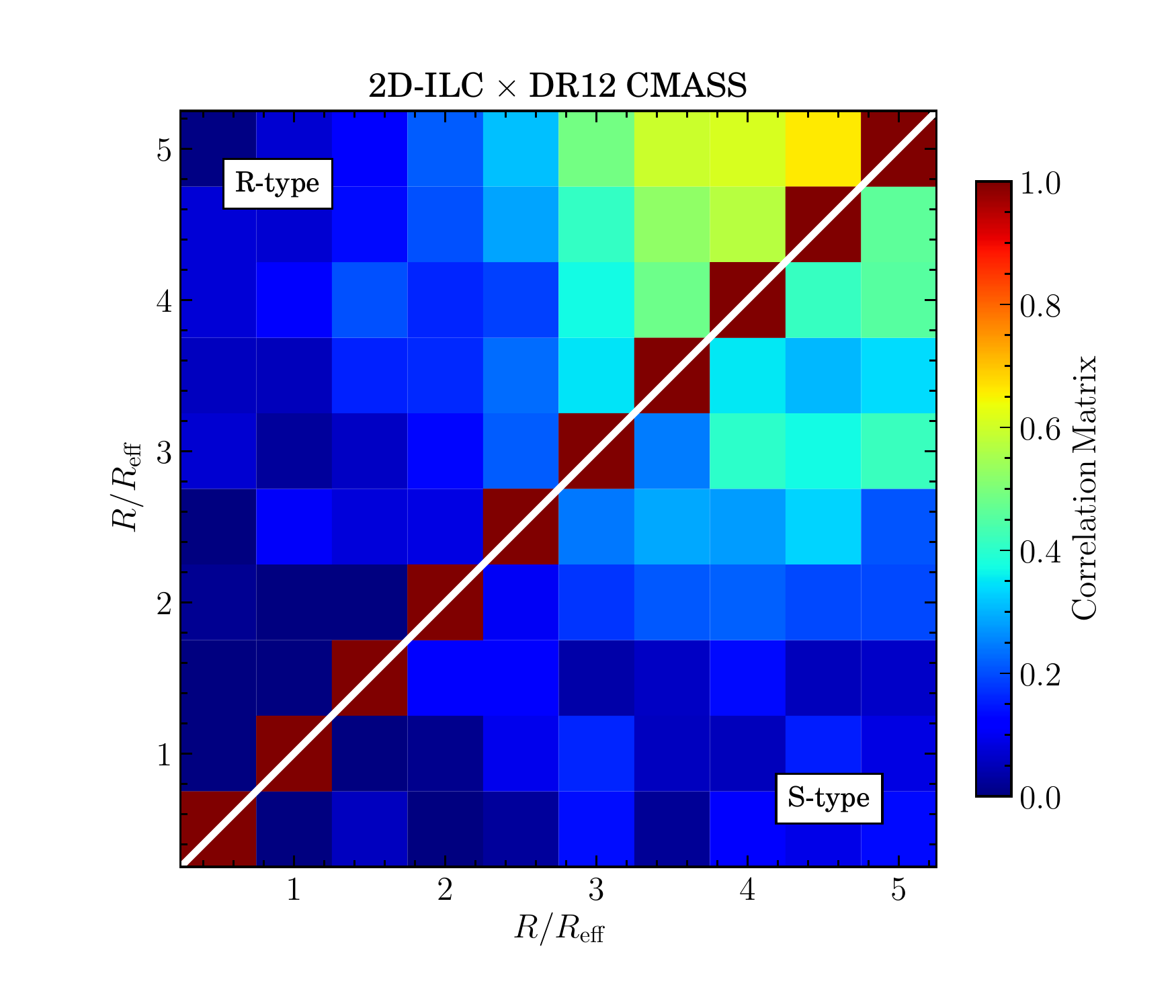}
    \caption{
        The correlation coefficient matrix for the measurements of
        \planck {\tt 2D-ILC} map with DR12 LOWZ/CMASS void catalogue
        (top-left/right panels, respectively); and 
        \planck {\tt SMICA-noSZ}/{\tt 2D-ILC} map with combined DR12 LOWZ and CMASS
        void catalogue (bottom-left/right panels, respectively).
        In each panel, the upper triangle shows the correlation coefficient
        matrix of the R-type voids; and lower triangle shows the S-type.
    }\label{fig:cov}
\end{figure*}

The results estimated with the \planck {\tt 2D-ILC} CMB map and 
LOWZ/CMASS galaxy and void catalogue are shown in \reffg{fig:dr12lv}.
The black step line is the same as in \reffg{fig:simlv}, which is
the theoretical prediction of the void expansion rate from simulation.

The error of the measurements is estimated by randomly-shuffling $N$ times the 
AP filtered CMB temperature values, which have galaxy samples located in.
The AP filter is applied before shuffling pixels. With such mock samples,
the voids relative position and involved galaxy number are kept the same,
but the LSS correlation are broken due to the shuffling.
The covariance matrix is then estimated as
\begin{align}
    C_{ij}^{\rm mock} = \frac{1}{N-1}\sum_{k=1}^{N} 
    \left(\hat{v}_k^{\rm e}(R_i) - \langle \hat{v}^{\rm e}(R_i)\rangle\right)
    \left(\hat{v}_k^{\rm e}(R_j) - \langle \hat{v}^{\rm e}(R_j)\rangle\right)
\end{align}
in which $i$ and $j$ are the indices of the $R/R_{\rm eff}$ bins, in which we have $10$
radius bins in this study.
$\hat{v}_k^{\rm e}(R)$ is the estimated radial velocity with the $k$-th
mock void catalogue realization;
$\langle\hat{v}^{\rm e}(R)\rangle$ is the mean of $N$ 
realizations and shown with the dashed lines in \reffg{fig:dr12lv}.
We use $N=300$ in our analysis and confirm that the covariance estimation is 
converged for $N\ge300$.
The diagonal terms of the covariance matrix are used as the error of the measurements, shown in Figs.~\ref{fig:dr12lv} and \ref{fig:simlv}.
The correlation coefficient matrix is then estimated via,
\begin{align}
    E_{ij} = \frac{C_{ij}^{\rm mock}}{\sqrt{C_{ii}^{\rm mock}C_{jj}^{\rm mock}}},
\end{align}
which is shown in \reffg{fig:cov}.
The correlation coefficient matrix for LOWZ and CMASS are shown in the left
and right panels, respectively. In each panel, the results for R-type voids
are shown in the upper triangle region; while the S-type voids are shown 
in the lower triangle region.
As we can see, at the larger $R/R_{\rm eff}$ bins, the results are more correlated 
between bins. The correlation between bins is due to the sharing of galaxy samples
between different voids. The galaxy samples located far from the void center have 
higher chance to be involved in different voids. Also we can see that the 
LOWZ samples are less correlated comparing to CMASS. It is because the 
CMASS samples are located at higher redshifts, and,
given the similar voids size distribution between CMASS and LOWZ,
galaxy samples at high redshift have more chance to be shared between voids.

The results for R-type voids ($\lambda_v<10$) are shown in the left panel.
The estimated kSZ temperature difference are converted to velocity by using \refeq{eq:T2v}.
The radial velocity profile for R-type voids is consistent with the
theoretical predictions within the errors. The fitted 
$\bar{\tau}=(1.24\pm0.79)\times10^{-4}$ and $(0.87\pm0.46)\times10^{-4}$ 
for LOWZ and CMASS, respectively, which are consistent with the results measured with 
pairwise kSZ of clusters~\cite{Li18} within the error.

To quantify the level of detection, we define $\chi^{2}_{\rm NULL}$ and 
$\chi^2_{\rm reduced}$ as follows
\begin{align}
    \chi^{2}_{\rm NULL} &= \sum_{ij} \hat{v}^{\rm e}(R_{i})
                          \left(C^{-1}\right)_{ij}
                          \hat{v}^{\rm e}(R_{j}), \label{eq:chisqnull} \\
    \chi^2_{\rm reduced} &= \chi^2_{\rm min} / {\rm d.o.f.} \label{eq:chisqredu}
\end{align}
where, ${\rm d.o.f.} = N_{\rm bins} - N_{\rm parameters}$, and
\begin{align}
    \chi^{2}_{\rm min} = {\rm min} \left[ 
    \sum_{ij}\left(\hat{v}^{\rm e}(R_{i})-v^{\rm t}(R_{i})\right)
    \left(C^{-1}\right)_{ij} 
    \left(\hat{v}^{\rm e}(R_{j}) - v^{\rm t}(R_{j}) \right) \right].
\end{align}

The $C^{-1}$ is the inverse of the covariance matrix. Besides the covariance 
matrix estimated with the mock void samples, we further include the covariance
of the intrinsic error $C^{\rm intr}$, which is due to the intrinsic scattering
of the velocity within the $R/R_{\rm eff}$ bins. 
The intrinsic covariance is estimated via velocities read directly from
the simulation and it contributes $\lesssim1\%$ of the uncertainties to the
total variance. But when the variance is reduced by using more
samples, the intrinsic variance contributes is up to $\lesssim10\%$.
Finally, we apply the Hartlap factor~\cite{2007A&A...464..399H}
to correct the biased inverse covariance matrix. The $C^{-1}$ is finally defined as
\begin{align}
    C^{-1} = \frac{N - N_{\rm bin} - 2}{N - 1} 
    \left(C^{\rm mock} + C^{\rm intr}\right)^{-1}.
\end{align}

Equation~(\ref{eq:chisqnull}) quantifies the detection of the signal with respect to null, 
where indices $i$ and $j$ run over all radius bins; and 
\refeq{eq:chisqredu} quantifies the goodness of fit.
We summarize our quantification in \reftb{tab:chi2-real}.

\subsubsection{Null detection}

In Figs.~\ref{fig:dr12lv} and \ref{fig:SMICAlv}, we plot the reconstructed peculiar velocities $\hat{v}^{\rm e}(R)$ with {\it Planck} {\tt 2D-ILC} and {\tt SMICA-noSZ} maps and LOWZ/CMASS void catalogues.  We list the detailed values of null detection for each respective case in Table~\ref{tab:chi2-real}. One can see that, for LOWZ catalogue, detection of R-type void is similar to the case of S-type void, and the $\chi^{2}_{\rm NULL}$ values are in the range of $7.11$ to $8.78$ depending on which CMB maps are used. One can further convert the $\chi^{2}_{\rm NULL}$ value into the ``$p$-value'', which is defined as the probability of no detection given the measurement, i.e.
\begin{eqnarray}
p={\rm Erfc}\left(\sqrt{\frac{\chi_{\rm NULL}^2}{2}}\right),
\end{eqnarray} 
where ``Erfc'' is the complementary error function. Therefore, for LOWZ catalogue, the probability of null detection of R-type and S-type voids for using two different versions of {\it Planck} CMB map is at $5\times 10^{-3}$ level, indicating $\sim 3\sigma$ C.L. detection of the two types of voids.

For CMASS catalogue, the R-type void is better detected than the S-type void, due to its uniformly expanding structure. The detection of R-type void is boosted to $\chi^{2}_{\rm NULL}=17.79$ for {\tt 2D-ILC} map and $14.69$ for {\tt SMICA-noSZ} map. This corresponds to the $p-$value as $2.47\times 10^{-5}$ for {\tt 2D-ILC} map and $1.27 \times 10^{-4}$ for {\tt SMICA-noSZ} map. For S-type void, the $\chi^{2}_{\rm NULL}$ is around $6.5$, which corresponds to $p=1.04 \times 10^{-2}$.

In Table~\ref{tab:chi2-real}, we also list the $\chi^{2}_{\rm NULL}$ for the combined LOWZ and CMASS catalogues. For R-type void, by using {\tt 2D-ILC} map it reaches $21.86$, and by using {\tt SMICA-noSZ} map, it reaches $21.32$. These correspond to the $p-$values in the range of $(2.93-3.89)\times 10^{-6}$. The S-type void detection is reduced to $\chi^{2}_{\rm NULL}\simeq 6.1$ for the {\tt 2D-ILC} map, for which $p-$value is $1.35\times 10^{-2}$.

To summarize, the probability of null detection for R-type and S-type voids are $2.93 \times 10^{-6}$ and $1.35 \times 10^{-2}$ for the {\it Planck} {\tt 2D-ILC} map, so the R-type and S-type voids are detected at $3.31\sigma$ and $1.75\sigma$ C.L. respectively.

\subsubsection{Measuring expansion profile}

For R-type voids, {\it Planck} {\tt 2D-ILC} map gives $\chi^2_{\rm reduced}=(0.33\,,1.34\,,1.19)$ with the
DR12-LOWZ, DR12-CMASS and the combination of such two voids 
catalogue, respectively. 
The lower $\chi^2_{\rm reduced}$ indicates that the results is over-fited.
It might be because the signal is attenuated due to the smear effect of the 
large AP filter. We have $1.54\sigma$ (weak) detection of the $\bar{\tau}$
with LOWZ cataluge, $1.94\sigma$ with CMASS catalogue and $2.68\sigma$
with the combination of such two catalogues.

For S-type voids, 
the reconstructed velocity profile is biased and the fitted $\bar{\tau}$ is 
consistent with $0$. 
This large uncertainty of estimates is due to the instrumental noise, 
CMB residual and other residual foreground. 
The estimated results are shown in right panel of \reffg{fig:dr12lv}.
The estimated kSZ temperature difference is also converted to velocity
using \refeq{eq:T2v}. However, we use the fitted $\bar{\tau}$ with R-type voids, 
instead of the S-type voids. 
We also did the estimation by 
replacing the {\tt 2D-ILC} map with {\tt SMICA-noSZ} map and find very similar 
results to Fig.~\ref{fig:dr12lv}. We then combine the DR12 LOWZ and CMASS samples
and stack the radial velocity profile against the {\tt 2D-ILC} and {\tt SMICA-noSZ}
maps, and show in Fig.~\ref{fig:SMICAlv}. One can see that the error-bars for 
both cases shrink and there is a clear detected signal for R-type of voids. 
For S-type of voids, there is still a large bias existed, due to its
``cross-zero line'' structure.

We then measure the value of optical depth ($\bar{\tau}$) by fitting the theoretical void expansion profile (black step line) to the data, and show our results in Table~\ref{tab:chi2-real}. The $\bar{\tau}$ value is measured to be $(1.03 \pm 0.38)\times 10^{-4}$ for the R-type void, and $(-0.17 \pm 0.24)\times 10^{-4}$ for the S-type void.

\begin{table*}
    {\scriptsize
    \centering
    \caption{
        The summary of the significance and bias of the reconstructed
        radial velocity profile of voids with {\it Planck} data.}\label{tab:chi2-real}
    \begin{tabular}{l|c|c|c|c|c|c|c|c|c|c} \hline 
        \multicolumn{5}{c|}{\multirow{2}{*}{LSS Tracer}} &
        \multicolumn{6}{c }{{\it Planck} maps} \\ \cline{6-11}

        \multicolumn{5}{c|}{} & 
        \multicolumn{3}{c|}{{\tt 2D-ILC}} &
        \multicolumn{3}{c }{{\tt SMICA-noSZ}} \\ \hline

        SDSS Catalogue              &
        $N_{\rm CG}$                &
        Effective Area              &
        Void type                   &
        $N_{\rm void}$              &
        $\chi^{2}_{\rm NULL}$       &
        $\chi^2_{\rm reduced}$      &
        $\bar{\tau}\times10^{4}$ &
        $\chi^2_{\rm NULL}$         &
        $\chi^2_{\rm reduced}$      &
        $\bar{\tau}\times10^{4}$ \\ \hline

        \multirow{2}{*}{DR12-LOWZ} &
        \multirow{2}{*}{$386,603$} &
        \multirow{2}{*}{$7,943\deg^2$} &
            R-type  & $1,462$ & $ 7.11$ & $ 0.33$ & $ 1.24\pm0.79$ 
                              & $ 7.68$ & $ 0.35$ & $ 1.30\pm0.77$\\ 
        &&& S-type  & $1,348$ & $ 8.78$ & $ 0.98$ & $-0.09\pm0.41$
                              & $ 7.71$ & $ 0.85$ & $-0.21\pm0.38$\\ \hline 

        \multirow{2}{*}{DR12-CMASS} &
        \multirow{2}{*}{$779,070$} &
        \multirow{2}{*}{$8,901\deg^2$} &
            R-type  & $3,845$ & $17.79$ & $ 1.34$ & $ 0.87\pm0.46$
                              & $14.69$ & $ 1.34$ & $ 0.59\pm0.45$\\ 
        &&& S-type  & $3,122$ & $ 6.57$ & $ 0.70$ & $-0.12\pm0.29$
                              & $ 6.96$ & $ 0.72$ & $-0.17\pm0.29$\\ \hline 

        \multirow{2}{*}{Combined} &
        \multirow{2}{*}{$1,165,673$} &
        \multirow{2}{*}{$8,901\deg^2$} &
            R-type  & $5,307$ & $21.86$ & $ 1.19$ & $ 1.03\pm0.38$
                              & $21.32$ & $ 1.51$ & $ 0.86\pm0.38$\\ 
        &&& S-type  & $4,470$ & $ 6.10$ & $ 0.62$ & $-0.17\pm0.24$
                              & $ 4.86$ & $ 0.45$ & $-0.22\pm0.24$\\ \hline 
    \end{tabular}
    }
\end{table*}

\section{Forecast for future CMB experiments}
\label{sec:forecasts}

\subsection{CMB residuals}

\begin{figure*}
    \small
    \centering
    \begin{minipage}[t]{0.49\textwidth}
        \includegraphics[width=\textwidth]{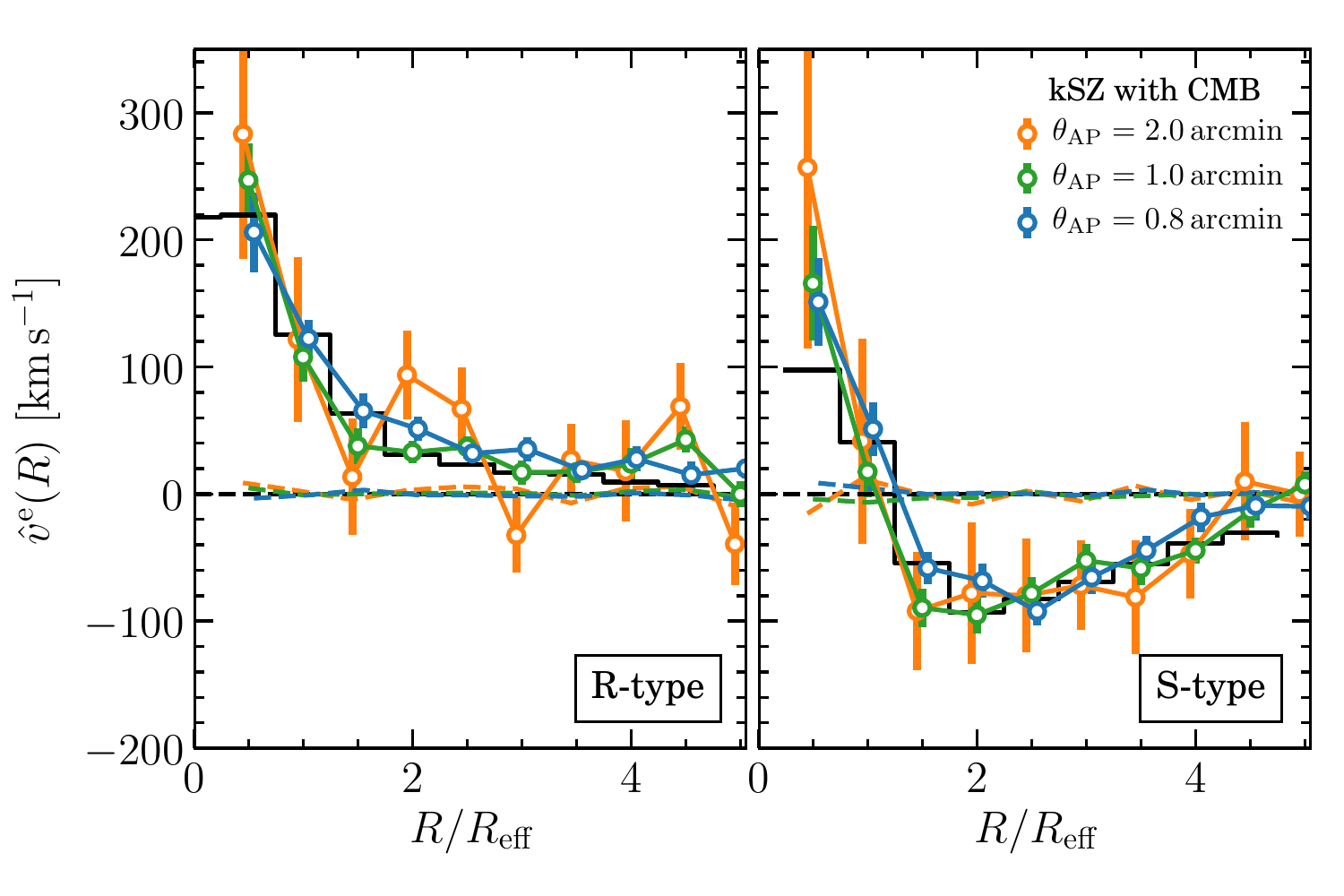}
        \caption{
            The simulated signal with added primary CMB as contaminations,
            assuming $2000\deg^2$ survey area and $\theta_{\rm FWHM}=1.4\,{\rm arcmin}$ 
            resolution.
            The estimated $\hat{v}^{\rm e}(R)$ with different AP filter radii
            are shown with different colors.
        }\label{fig:simcmb_ap}
    \end{minipage}\hfill
    \begin{minipage}[t]{0.49\textwidth}
        \includegraphics[width=\textwidth]{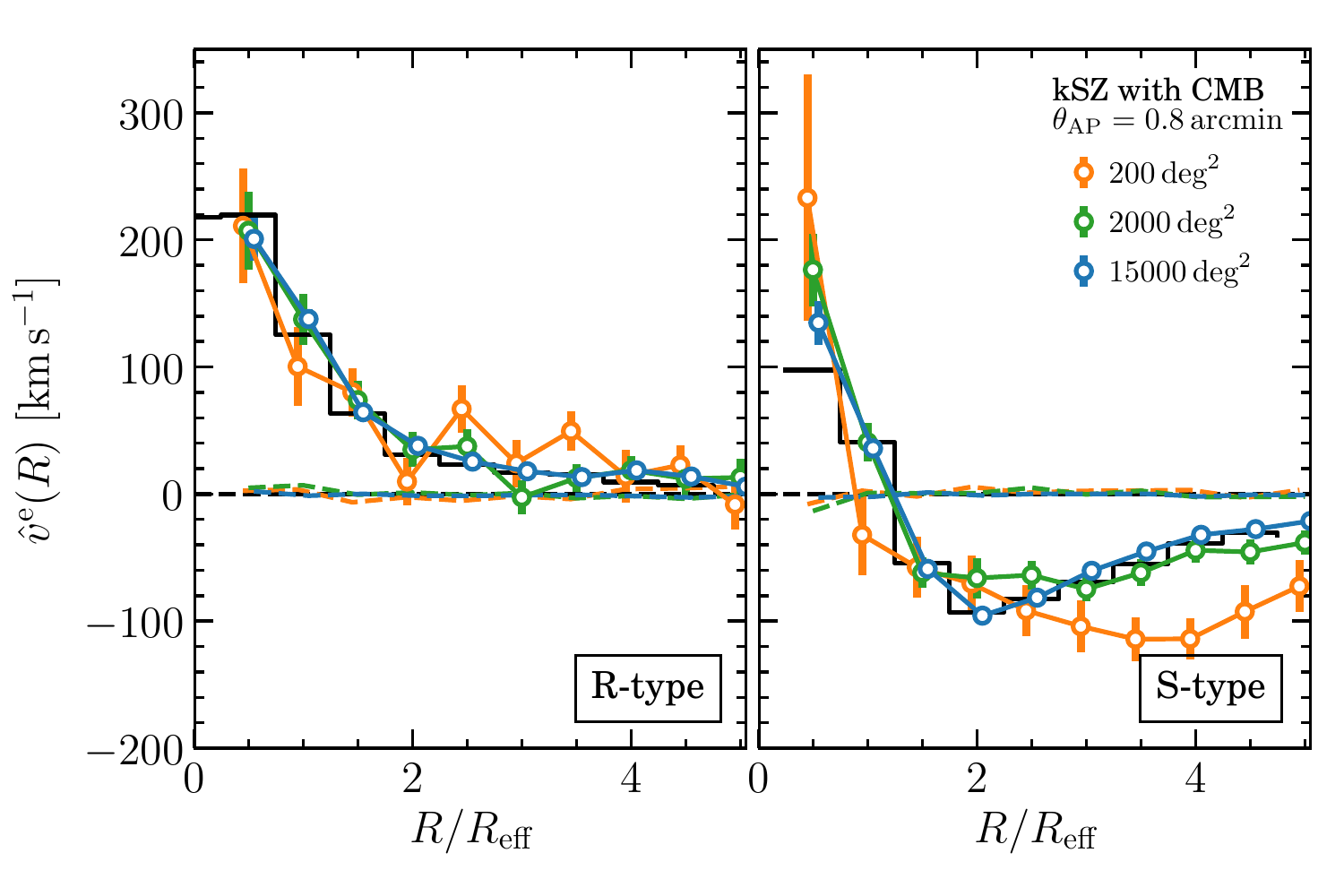}
        \caption{
            The simulated signal with CMB fluctuation residual as contaminations,
            assuming $\theta_{\rm FWHM}=1.4\,{\rm arcmin}$ resolution and 
            $\theta_{\rm AP}=0.8\,{\rm arcmin}$ as filter
            radius. Different color indicate the results with different survey area.
        }\label{fig:simcmb_area}
    \end{minipage}
\end{figure*}

The major contaminations to the signal come from the CMB temperature fluctuations,
which are supposed to be filtered out with the designed filter.
However, the residual of the CMB temperature fluctuation is still not negligible
comparing to the relatively weak signal. In the meanwhile, the residual is also filter-parameter
dependent. In this analysis, we focus on discussing the effect of the AP filter.

We add the simulated CMB temperature fluctuations to the signal map, which has
resolution of $1.4\,{\rm armin}$ and the area of $2000\,{\rm deg}^2$. 
The AP filter is applied with filter radius sizes varying between $2\,{\rm arcmin}$,
$1\,{\rm arcmin}$ and $0.8\,{\rm arcmin}$. The results are shown in
\reffg{fig:simcmb_ap}. As shown in the figure, the measured $\tilde{v}^{\rm e}_R$ 
with $2\,{\rm arcmin}$ AP radius size (shown in orange color) has large errors.
The errors are reduced with smaller AP filter radius size.
With $2\,{\rm arcmin}$, $1\,{\rm arcmin}$ and $0.8\,{\rm arcmin}$ AP size,
we achieve 
$2.93\sigma\,\,,  7.68\sigma\,\,, 9.11\sigma$ detection for R-type voids; and
$3.63\sigma\,\,, 11.82\sigma\,\,,11.40\sigma$ for S-type.
The corresponding fitted $\bar{\tau}$ are 
$(2.26 \pm 0.77\,\,, 2.95 \pm 0.38\,\,, 3.51 \pm 0.38)\times 10^{-4}$
for R-type voids ($\lambda_v<10$) and 
$(2.88 \pm 0.79\,\,, 5.12 \pm 0.43\,\,, 4.66 \pm 0.41)\times 10^{-4}$
for S-type ($\lambda_v>10$). 
The smaller value of $\bar{\tau}$ with larger AP radius 
size is due to the smear effect of the AP filter.

In the mean while, an obvious bias is shown at some scales, especially 
with large AP filter size.
With different simulation realizations, we found that the amplitude and 
scales of the bias are varying. A possible reason of the bias is
the residual structures of the CMB temperature fluctuation. 
The larger the AP filter size, the more residual CMB temperature fluctuation
are existed in the estimation.
On the other hand, the bias is also dependent on the number of samples.
To investigate how the bias are varying with voids sample, we use 
different effective area of the simulated catalog, assuming CMB map
resolution of $1.4\,{\rm armin}$ and AP filter size of $0.8\,{\rm arcmin}$.
The results are shown in \reffg{fig:simcmb_area}. 
The green data in \reffg{fig:simcmb_area}
shows the results assuming $2,000\deg^2$ survey area, 
which includes $410$ R-type and $174$ S-type voids; 
The orange ones are the results by selecting only 
$10\%$ of the survey area as the green ones, corresponding to reduce
the survey area to $200\deg^2$ and including $36$ R-type and $20$ S-type voids;
The blue data are the results with
$15,000\deg^2$, including $3,002$ R-type and $1,403$ S-type voids.
With the survey area varied from $200\deg^2$, $2,000\deg^2$ to $15,000\deg^2$,
the $\chi^2_{\rm reduced}$ vary from $1.81$, $0.98$ to $1.11$ for the R-type voids,
and from $3.94$, $1.47$ to $0.98$ for the S-type voids.
Given the ${\rm d.o.f}=9$, a $\chi^2_{\rm reduced} = 1.00 \pm 0.47$ indicates 
a good fit.

\subsection{Future experiments}

\begin{table*}
    {\scriptsize
    \centering
    \caption{
        Forecasts for ACT, AdvACT and Simons Observatory 
        detection of voids profile. 
        The results with $\sigma_{n}=0$ is the instrumental noise free case.
    }\label{tab:chi2}
    \begin{tabular}{C{1.5cm}|C{1.5cm}|C{1.5cm}|c|c|c|c|c|l} \hline 
        Survey Area [deg$^{2}$]
        & $N_{\rm CG}$
        & $\sigma_{\rm n}$ [$\mu$K-arcmin] 
        & Void type
        & $N_{\rm void}$ 
        & $\chi^{2}_{\rm NULL}$ 
        & $\chi^2_{\rm reduced}$ 
        & $\bar{\tau}\times10^{4}$
        & Experiments \\ \hline

        \multirow{4}{*}{$2,000 $}  & \multirow{4}{*}{$50,250$} &
             \multirow{2}{*}{$ 0$} &
                    R-type & $410$ & $147.91$ & $0.98$ & $3.51\pm0.38$ &\multirow{4}{*}{SPT-3G} \\
          &&      & S-type & $174$ & $119.77$ & $1.47$ & $4.66\pm0.41$ & \\ \cline{3-8}
          && \multirow{2}{*}{$10$} &
                    R-type & $410$ & $ 45.46$ & $0.94$ & $4.30\pm1.18$ &\\
          &&      & S-type & $174$ & $ 26.68$ & $1.42$ & $5.38\pm1.27$ &\\ \hline
        \multirow{6}{*}{$15,000 $} & \multirow{6}{*}{$394,155$} &
             \multirow{2}{*}{$20$} &
                    R-type & $3002$& $ 26.79$ & $0.84$ & $3.03\pm0.89$ &\multirow{2}{*}{Pessimistic AdvACT}\\
          &&      & S-type & $1403$& $ 36.76$ & $1.65$ & $3.87\pm0.82$ &\\ \cline{3-9}
          && \multirow{2}{*}{$10$} &
                    R-type & $3002$& $ 97.43$ & $1.39$ & $3.60\pm0.51$ &\multirow{2}{*}{Simons O./AdvACT}\\
          &&      & S-type & $1403$& $ 47.05$ & $0.82$ & $3.22\pm0.46$ &\\ \cline{3-9}
          && \multirow{2}{*}{$ 6$} &
                    R-type & $3002$& $204.49$ & $1.03$ & $3.65\pm0.34$ &\multirow{2}{*}{Optimistic Simons O./AdvACT}\\
          &&      & S-type & $1403$& $133.48$ & $1.89$ & $3.68\pm0.31$ &\\ \hline
    \end{tabular}%
    }
\end{table*}

\begin{figure*}
    \small
    \centering
    \begin{minipage}[t]{0.49\textwidth}
        \includegraphics[width=\textwidth]{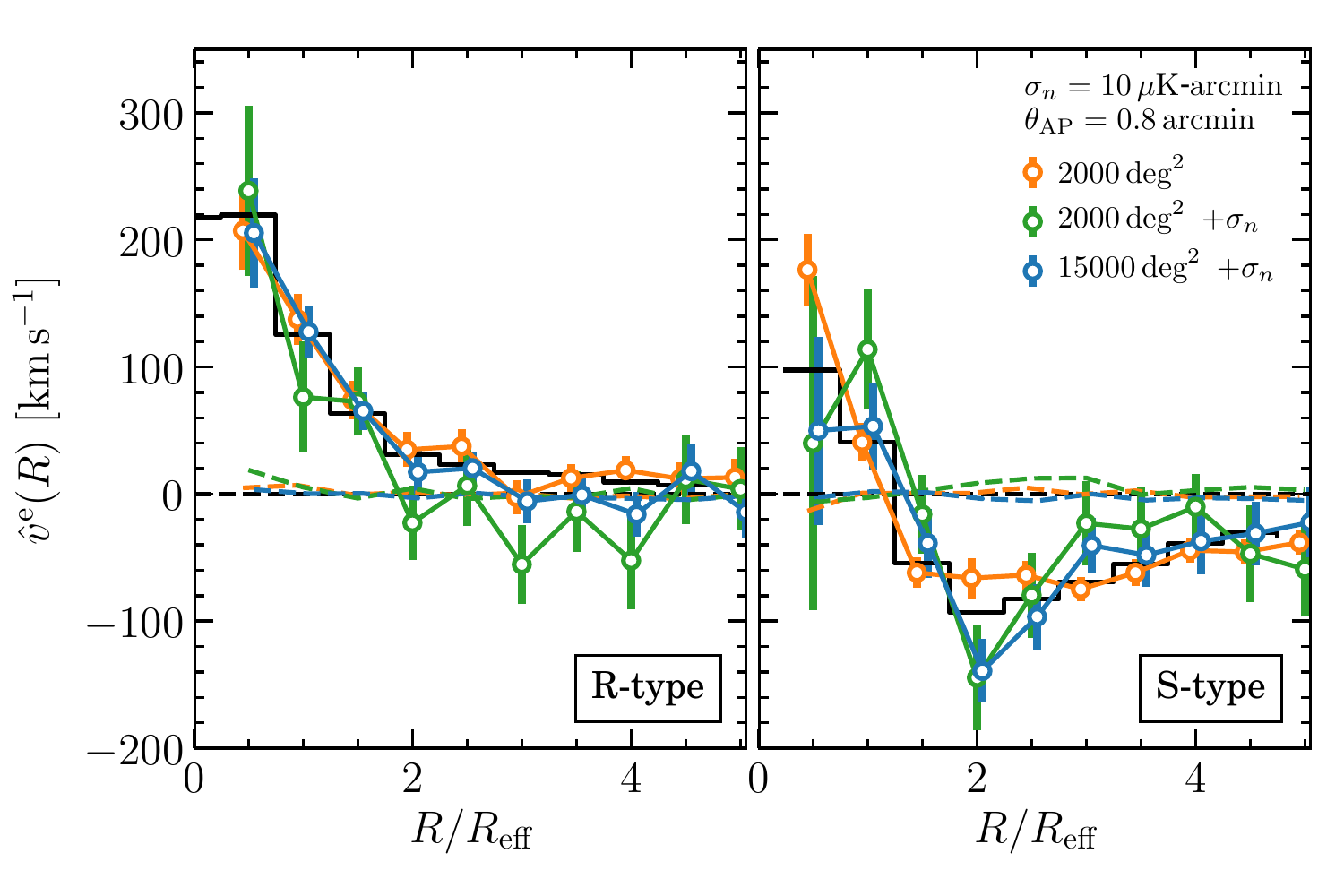}
        \caption{
            The simulated signal with CMB fluctuation residual and thermal noise
            of $10{\rm \mu K \minus arcmin}$, assuming $\theta_{\rm FWHM}=1.4\,{\rm arcmin}$ 
            resolution and $\theta_{\rm AP}=0.8\,{\rm arcmin}$ AP filter radius. 
            The result with $2,000\deg^2$ and $15,000\deg^2$ survey area are
            shown with different colors.}\label{fig:simcmbn10}
    \end{minipage}\hfill
    \begin{minipage}[t]{0.49\textwidth}
        \includegraphics[width=\textwidth]{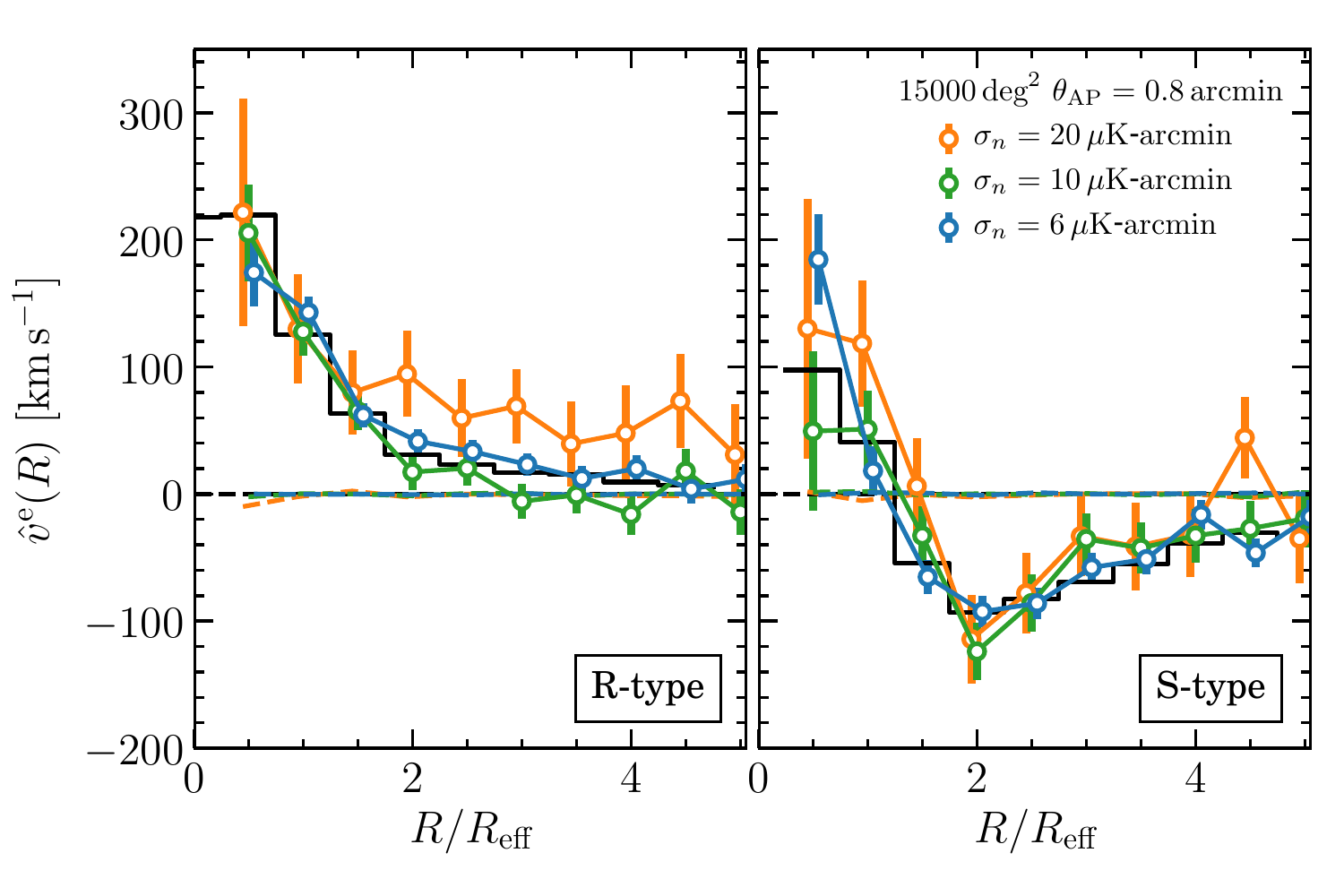}
        \caption{
            The simulated signal with CMB fluctuation residual and thermal noise,
            assuming $\theta_{\rm FWHM}=1.4\,{\rm arcmin}$ resolution and $\theta_{\rm AP}=0.8\,{\rm arcmin}$ as filter
            radius. Different colors indicate the results with different thermal
            noise level.
        }\label{fig:simcmbarea}
    \end{minipage}
\end{figure*}

The sensitivity and angular resolution of CMB experiments are being improved rapidly, with several on-going ground-based CMB observatories, such as
SPT-3G~\cite{Benson14}, ACT~\cite{Hilton18}, AdvACT
\cite{2014SPIE.9153E..1PB,2016JLTP..184..805S,2016JLTP..184..772H}, 
and Simons Observatory~\cite{Simons-2019JCAP}. These ground-based CMB surveys will
provide higher angular resolution CMB maps with lower noise levels in the near future. 
Here we plan to simulate the future detectability of ACT, AdvACT and Simons Observatory, 
by assuming that the angular resolution of these survey can reach 
$\theta_{\rm FWHM}=1.4\,$arcmin. 
We assume that SPT-3G will cover $2,000\,{\rm deg}^{2}$ survey area, 
AdvACT and Simons Observatory will cover $15,000\,{\rm deg}^{2}$ area,
respectively~\cite{Simons-2019JCAP}. 
The latter one corresponds to the $f_{\rm sky}\simeq 0.4$ but excluding Galactic plane. 
Beside the CMB temperature fluctuation, we further add instrumental noises at 
different level according to different CMB experiments.
We assume $10\,\mu{\rm K}\minus$arcmin {\it rms} noise for SPT-3G. 
For AdvACT, the {\it rms} noise level is around 10 to 20 $\mu{\rm K}\minus$arcmin so we 
assume that $\sigma_{\rm n}=20\,\mu{\rm K}\minus$arcmin as pessimistic case, 
and $\sigma_{\rm n}=10\,\mu{\rm K}-$arcmin as optimistic case. 
For Simons Observatory, we assume $\sigma_{\rm n}=10\,\mu{\rm K}\minus$arcmin 
as pessimistic case and $\sigma_{\rm n}=6\,\mu{\rm K}\minus$arcmin as 
optimistic case as indicated in Ref.~\cite{Simons-2019JCAP}.
The results are shown in Figs.~\ref{fig:simcmbn10} and \ref{fig:simcmbarea}; and 
the fitting values are summarized in \reftb{tab:chi2}.

Figure~\ref{fig:simcmbn10} shows the results with {\it rms}
$=10\,\mu{\rm K}\minus{\rm arcmin}$ noise added. 
The result with noise free is shown in orange color
and the result with $2,000\deg^2$ survey area is shown in green.
Although the estimated $\bar{\tau}$ is consistent with each other,
the error is increased from $0.38 (0.41)\times10^{-4}$ to $1.18 (1.27)\times10^{-4}$
for R-type (S-type) voids. Even though the error is increased by a factor of $3$,
the further CMB experiments such as SPT-3G can still achieve 
$3.65\sigma (4.22\sigma)$ detection with $2,000\deg^2$ survey area.

If the survey area can be extended to $15,000\deg^2$, the estimation uncertainty
can be reduced significantly. In the case of pessimistic case of AdvACT experiments,
one can still achieve $3.4\sigma (4.73\sigma)$ detection for R-type (S-type) voids with noise 
{\it rms}$=20\,\mu{\rm K}\minus{\rm arcmin}$. With the optimistic AdvACT experiment
or the pessimistic Simons Observatory, by assuming noise {\it rms}$=10\,\mu{\rm K}\minus{\rm arcmin}$, the detection can be improved to 
about $7\sigma$ for both R-type and S-type voids. With further optimistic case of
noise {\it rms}$=6\,\mu{\rm K}\minus{\rm arcmin}$ for Simons Observatory, one can potentially achieve $\sim 10.7\sigma$ (R-type) and $11.8\sigma$ (S-type) detections with our estimator.

\section{Conclusion}\label{sec:conclusion}
In this paper, we develop an estimator to extract the mean radial velocity 
profile of the voids via the kinematic Sunyaev-Zeldovich effect of pairs of galaxies around the voids.
We firstly test the estimator with the simulated pure kSZ maps, which are
constructed from the BigMD $N$-body simulation 
\cite{2016MNRAS.457.4340K,2012MNRAS.423.3018P} and the void catalogue
generated with the same simulation. The estimator is tested with the R-type and 
S-type voids separately and the results show that the mean radial velocity 
profile can be fully recovered with our estimator. With the simulated kSZ map, we note that the recovered signal is highly
attenuated with the lower angular resolution of the map or larger AP filter
radius size, leading to the potential bias of optical depth.

We apply the estimator to the \planck {\tt 2D-ILC} CMB map, with the
LOWZ/CMASS galaxy and void catalogue. We conducted two statistics for the measured expansion profile of the void. One is the $\chi_{\rm NULL}^{2}$ for NULL detection, which is calculated as the data against the zero line; the other is the $\chi^{2}_{\rm reduced}$ which is calculated as the minimal $\chi^{2}$ value between the data and the best-fitting theoretical profile. For the R-type and S-type voids, $\chi^{2}_{\rm NULL}$ is calculated as $21.86$ and $6.10$ for the {\tt 2D-ILC} map, which corresponds to the probability of $2.93\times 10^{-6}$ and $1.35\times 10^{-2}$ of null detection. This suggests that the R-type and S-type voids are measured at $3.31\sigma$ and $1.75\sigma$ C.L. respectively.

We also calculate the $\chi^2_{\rm reduced}$ value as the minimal $\chi^{2}$ of the data with respect to the theoretical expansion profile. For R-type voids, the $\chi^2_{\rm reduced}=(0.33\,,1.34\,,1.19)$ with the
DR12-LOWZ, DR12-CMASS and the combination of such two voids 
catalogue, respectively. 
We have $1.54\sigma$ measurement of the optical depth $\bar{\tau}$
with LOWZ cataluge, $1.94\sigma$ with CMASS catalogue and $2.68\sigma$
with the combination of such two catalogue.
For the S-type voids, the measured $\bar{\tau}$ value is still consistent with $0$.

We further investigate the effect of the CMB temperature fluctuation contamination, 
by adding simulated CMB map to the kSZ simulation map. 
The AP filter is applied to the simulated map. The results show that
the CMB temperature fluctuation residual can increase the uncertainty of estimation and bias the results.
The uncertainty and bias can be reduced with smaller angular size of AP filter and larger amount
of voids samples.

Finally, we forecast the detection for a few future CMB experiments with higher angular resolution and lower thermal noise. Beside the CMB contamination, we further
add the instrumental noise at different level.
The forecasts show that, experiment like SPT-3G can obtain
$3.65\sigma (4.22\sigma)$ detection with $2,000\deg^2$ survey area at 
{\it rms}=$10\,\mu{\rm K}\minus{\rm arcmin}$; 
With $15,000\deg^2$ survey area, experiment like AdvACT can obtain
$3.4\sigma (4.73\sigma)$ detection for R-type (S-type) voids with noise 
{\it rms}$=20\,\mu{\rm K}\minus{\rm arcmin}$; or
about $7\sigma$ for both R-type and S-type voids at noise level of
{\it rms} $=10\,\mu{\rm K}\minus{\rm arcmin}$. Finally, in the case
of {\it rms} $=6\,\mu{\rm K}\minus{\rm arcmin}$, which is the 
most optimistic case of Simons Observatory, one can achieve over $10\sigma$
detection. Since the cosmic void is the most abundant structures of the large-scale Universe, our estimator opens a new window of probing dynamics of cosmic structures through the measurement of kinetic Sunyaev-Zeldovich effect.

\acknowledgements
We would like to thank Mathieu Remazeilles for supplying his {\it Planck} {\tt 2D-ILC} 
map for this work and helpful discussion, and the useful discussion with 
Matthew Hilton and Anthony Walters. YZM would like to acknowledge the supports 
from National Research Foundation with grant no. 105925, 109577, 120378, and 120385, and National Science Foundation China with grant no. 11828301.

\bibliography{draft}
\bibliographystyle{apsrev}

\end{document}